\documentclass[prl, twocolumn, floatfix, groupedaddress]{revtex4}
\usepackage{graphicx}
\usepackage{amsmath}
\usepackage{amssymb}
\usepackage{bm}
\usepackage{color}
\usepackage{hyperref}
\usepackage{tabularx}
\usepackage{multirow}

\begin{document}

\title{Superconductivity-Induced Ferromagnetism and Weyl Superconductivity in Nb-doped Bi$_2$Se$_3$}

\author{Noah F. Q. Yuan, Wen-Yu He, K. T. Law} \thanks{Correspondence address : phlaw@ust.hk}

\affiliation{$^1$Department of Physics, Hong Kong University of Science and Technology, Clear Water Bay, Hong Kong, China\\
$^2$Center for 2D and 1D Quantum Materials, Hong Kong University of Science and Technology, Clear Water Bay, Hong Kong, China}

\begin{abstract}
{A recent experiment on Nb-doped Bi$_2$Se$_3$ showed that zero field magnetization appears below the superconducting transition temperature. This gives evidence that the superconducting state breaks time-reversal symmetry spontaneously and is possibly in the chiral topological phase. This is in sharp contrast to the Cu-doped case which is possibly in the nematic phase and breaks rotational symmetry spontaneously. By deriving the free energy of the system from a microscopic model, we show that the magnetic moments of the Nb atoms can be polarized by the chiral Cooper pairs and enlarge the phase space of the chiral topological phase compared to the nematic phase. We further show that the chiral topological phase is a Weyl superconducting phase with bulk nodal points which are connected by surface Majorana arcs.}
\end{abstract}

\maketitle
{\bf Introduction}---
Superconductivity and ferromagnetism are conventionally considered to be two competing orders, especially in time-reversal invariant superconductors   \cite{Ginzburg, Berk}. However, superconductivity discovered in several Uranium based ferromagnetic compounds  \cite{UGe2, URhGe, UCoGe} indicates that the superconducting and ferromagnetic orders can also coexist. The muon spin rotation ($ \mu $SR) experiments in systems such as the B phase of UPt$_3$ \cite{UPt3}, Sr$_2$RuO$_4$  \cite{Sr2RuO4}, LaNiC$_2$  \cite{Hillier1}, LaNiGa$_2$ \cite{Hillier2}, SrPtAs  \cite{SrPtAs,TRSB-Sigrist} and URu$_2$Si$_2$ \cite{URu2Si2-Li,URu2Si2-Schemm} suggest that the time-reversal symmetry can be spontaneously broken in the superconducting phase. In this work, we propose that in Nb-doped bismuth selenide (Bi$_2$Se$_3$), ferromagnetism can be induced by chiral topological superconductivity and result in a three-dimensional Weyl topological superconductor.

Bi$_2$Se$_3$ are topological insulators  \cite{TI0, TI1, TI2, 3DTI} which support topological surface states. Soon after the topological surface states in Bi$_2$Se$_3$ were observed in angle-resolved photoemission (ARPES) experiments  \cite{TI3, TI4}, it was found that Cu-doped Bi$_2$Se$_3$ were superconducting  \cite{CuTI1, CuTI2, CuTI3}. It was proposed by Fu and Berg that these superconducting Cu-doped Bi$_2$Se$_3$ were in the odd-parity topological phase with Majorana surface states \cite{Fu1}. The Majorana surface states were supposed to induce zero bias conductance peaks in tunneling experiments  \cite{CuTI-Ando} but negative results were also found  \cite{CuTI-Chu}. Interestingly, the Knight shift experiments  \cite{GQZheng} showed that spin-susceptibility of the material has a two-fold in-plane rotational symmetry which breaks the three-fold rotational symmetry of the basal plane. This motivated Fu to propose that the superconducting Cu-doped Bi$_2$Se$_3$ is in the so-called nematic phase which spontaneously breaks the three-fold rotational symmetry down to a two-fold symmetry  \cite{Fu2}. In a more recent measurement, the two-fold dependence of the specific heat on the applied in-plane magnetic field was found which also supports the superconducting nematic phase scenario \cite{Ando}.

\begin{figure}
\includegraphics[width=3.50in]{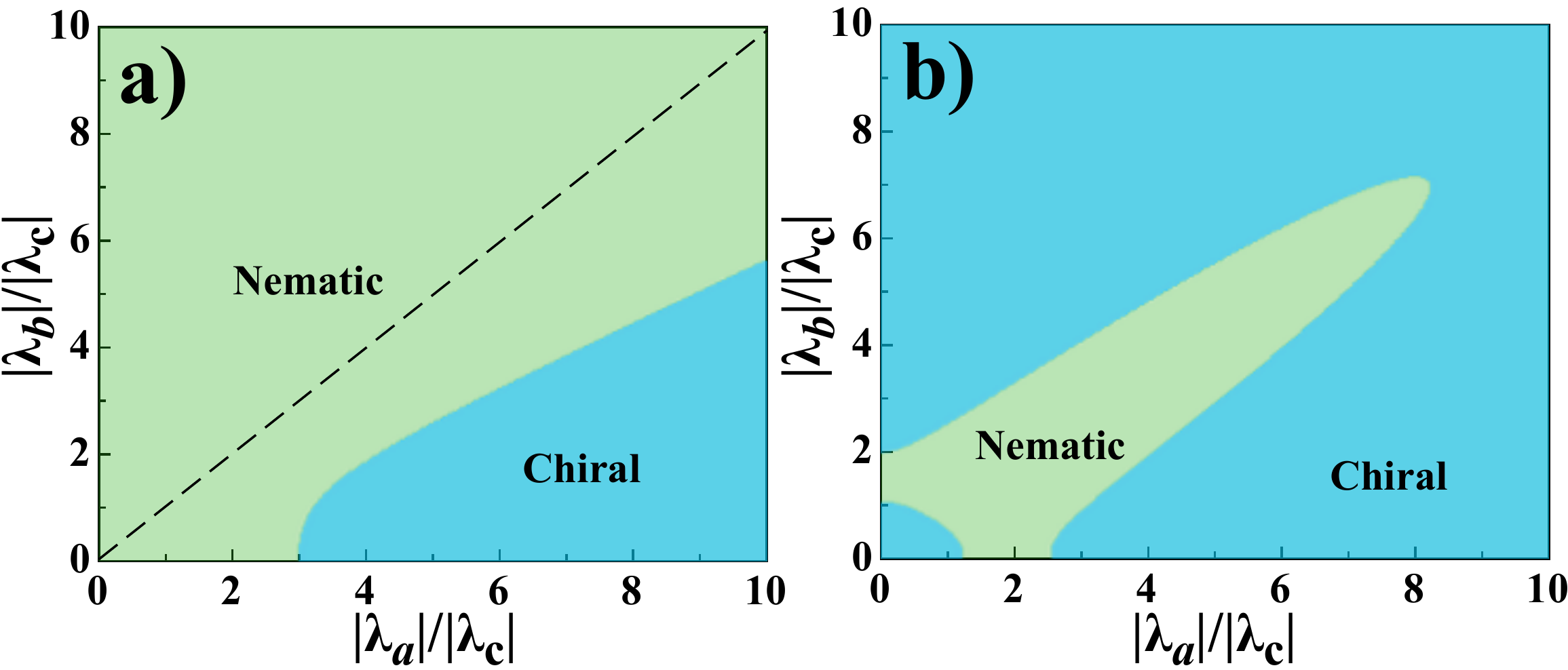}
\caption{Phase diagrams of doped Bi$_2$Se$_3$, the blue region denotes the chiral phase while the green region denotes the nematic phase. a) The case for $J\mu_d = 0$ which is relevant to the Cu-doped case when $\mu_d=0$. If only the $\Gamma_0$ terms in Eqs. (\ref{Pairing+}) and (\ref{Pairing-}) are finite as assumed in Ref. \cite{Fu3}, the system is always in the nematic phase with $\lambda_a  = -\lambda_b, \lambda_c =0$ as shown by the dashed line. b) The case for finite $J \mu_d$ which is relevant to the Nb-doped case.  The chiral phase regime is enlarged. }
\label{fg2}
\end{figure} 

The superconducting nematic phase belongs to the $E_{u}$ representation of the $D_{3d}$ point group  \cite{Fu1}. Interestingly, the two-dimensional $E_{u}$ representation also allows the appearance of the chiral topological phase which breaks time-reversal symmetry spontaneously  \cite{Fu3}. Recently, in the experiment performed by Qiu \textit{et al.} on Nb-doped Bi$_2$Se$_3$, it was found that the sample has finite magnetizations at zero field below the superconducting transition temperature ($T_c$)  \cite{NbTI1}. The spontaneous time-reversal symmetry breaking below $T_c$ suggests that superconducting Nb-doped Bi$_2$Se$_3$ is possibly in the chiral topological phase.

In this work, using a microscopic Hamiltonian which includes the coupling between the itinerant electrons of Bi$_2$Se$_3$ and the magnetic moments of Nb, we derive the mean field free energy of the system. We show that the superconducting order parameters of the itinerant electrons couple to the ferromagnetic order parameter of the Nb atoms. The chiral superconducting phase induces a finite ferromagnetic order parameter, whereas the nematic phase cannot induce any ferromagnetic order. Moreover, the superconductivity-ferromagnetism coupling enlarges the phase space of the chiral superconducting phase as depicted in Fig.1. This provides a possible explanation of why the Cu-doped and Nb-doped Bi$_2$Se$_3$ have different superconducting states. We further show that the chiral topological superconductor is a Weyl superconductor with bulk nodal points and surface Majorana arcs.

In the following sections, we first propose the model Hamiltonian for both Cu-doped and Nb-doped Bi$_2$Se$_3$ and derive the mean field free energy. Second, by minimizing the Ginzburg-Landau (GL)  free energy, we show that the induced magnetization by the chiral superconducting phase is nonzero. Third, the phase diagram in the presence of dopants with finite magnetic moments is studied. Finally, the energy spectrum of the chiral topological phase is analyzed. 

{\bf Model Hamiltonian}---
Experimentally, the normal phases of both Cu-doped and Nb-doped Bi$_2$Se$_3$ are found to be paramagnetic \cite{CuTI1, NbTI1}. Hence, the normal state symmetry group of the doped Bi$_2$Se$_3$ can be chosen as $ D_{3d}\times\mathbb{T} $ where $ \mathbb{T} $ is the time-reversal symmetry.  For the  itinerant electrons in doped Bi$_2$Se$_3$ near the Fermi energy, the dominant orbitals are the $p$-orbitals from both Bi and Se atoms. To the first order in momentum $\bm k$, the Hamiltonian takes the form \cite{TI1, TI2, Fu1, Fu3}: 
\begin{equation}
H_{e}=\sum_{\bm k}c^{\dagger}_{\bm k}[ v(k_y\sigma_x -k_x\sigma_y)\tau_z +v_{z}k_{z}\tau_y +m\tau_x -\mu ]c_{\bm k},
\end{equation}
where $ c_{\bm k}=\{c_{\bm k a\uparrow},c_{\bm k a\downarrow},c_{\bm k b\uparrow},c_{\bm k b\downarrow}\}^{\text{T}} $ consists of the $a$ and $b$ orbitals originating from the $p$-orbitals of the Bi and Se atoms and $\uparrow, \downarrow$ denote the $z$-component of the total angular momentum \cite{SM}. The Pauli matrices $ \bm\sigma =\{\sigma_x ,\sigma_y ,\sigma_z\} $ and $ \bm\tau =\{\tau_x ,\tau_y ,\tau_z\} $ act on the spin space and the orbital space respectively. The doped chemical potential $ \mu $ lies within the conduction band \cite{NbTI1, CuTI2}.

To consider the effect of dopant atoms, we ignore the dopant-dopant interactions as the normal phase is paramagnetic. The electron-dopant coupling is assumed to be short-ranged and can be written as:
\begin{equation}\label{ed}
H_{ed}=\sum_{\bm r}[I\sum_{i=x,y}s_{i}(\bm r)m_{i}(\bm r)
+Js_{z}(\bm r)m_{z}(\bm r)].
\end{equation}
Here, the operator $ \bm s(\bm r)=c^{\dagger}(\bm r)\bm\sigma c(\bm r) $ denotes the total spin density of electrons, while $ \bm m(\bm r) $ denotes the magnetic moment density of dopant atoms. The constants $ I$ and $J$ are the magnetic exchange coupling strengths. It is important to note that $H_{ed}$ respects the $ D_{3d}\times\mathbb{T} $ symmetry and the symmetry properties of the operators are shown in Table.\ref{G}.

Motivated by recent experiments in which point group symmetry and time-reversal symmetry are spontaneously broken in Cu-doped and Nb-doped Bi$_2$Se$_3$ respectively, we assume that the superconducting phases belong to the two-dimensional representation of the $D_{3d}$ point group. To be specific, we consider the odd-parity $ E_{u} $ representation of $D_{3d}$ \cite{Fu1, Fu2, Fu3}. In this $E_{u}$ phase, there are two kinds of Cooper pairs which are related by time-reversal symmetry. These two kinds of chiral Cooper pairs are created by $F_{+}^{\dagger},F_{-}^{\dagger}$ respectively, which to the first order in $ \bm k $ can be written as:
\begin{eqnarray}\nonumber
F_{+}^{\dagger}&=&\sum_{\bm k}\left[4i\Gamma_{0}c_{\bm k a\uparrow}^{\dagger}c_{-\bm k b\uparrow}^{\dagger}+\Gamma  k_{+}\left(c_{\bm k a\uparrow}^{\dagger}c_{-\bm k b\downarrow}^{\dagger}+c_{\bm k a\downarrow}^{\dagger}c_{-\bm k b\uparrow}^{\dagger}\right)\right.  \\\label{Pairing+}
&+&2\left.\sum_{l=a,b}\left(i\Gamma_{xy}k_{-}c_{\bm k l\downarrow}^{\dagger}c_{-\bm k l\downarrow}^{\dagger}+\Gamma_{z}k_{z}c_{\bm k l\uparrow}^{\dagger}c_{-\bm k l\uparrow}^{\dagger}\right)\right],\\\nonumber
F_{-}^{\dagger}&=&\sum_{\bm k}\left[-4i\Gamma_{0}c_{\bm k a\downarrow}^{\dagger}c_{-\bm k b\downarrow}^{\dagger}+\Gamma  k_{-}\left(c_{\bm k a\uparrow}^{\dagger}c_{-\bm k b\downarrow}^{\dagger}+c_{\bm k a\downarrow}^{\dagger}c_{-\bm k b\uparrow}^{\dagger}\right)\right.  \\\label{Pairing-}
&+&2\left.\sum_{l=a,b}\left(i\Gamma_{xy}k_{+}c_{\bm k l\uparrow}^{\dagger}c_{-\bm k l\uparrow}^{\dagger}-\Gamma_{z}k_{z}c_{\bm k l\downarrow}^{\dagger}c_{-\bm k l\downarrow}^{\dagger}\right)\right].
\end{eqnarray}
Here, $k_{\pm}=k_x \pm i k_y$. The $\Gamma_0$ terms create Cooper pairs formed by electrons with different orbitals and equal spins, the $\Gamma$ terms create inter-orbital $p \pm ip$-wave Cooper pairs. The $\Gamma_{xy}$ and $\Gamma_{z}$ terms create Cooper pairs formed by electrons with the same orbital and with $p \pm ip$ and $p_z$ pairings respectively. The creation operator $F_{-}^{\dagger}$ is the time-reversal partner of $F_{+}^{\dagger}$ and the two operators form the $E_{u}$ representation of $D_{3d}$.

Thus, the phenomenological electron-electron interaction Hamiltonian of the doped Bi$_2$Se$_3$ can be expressed in terms of Cooper pair creation operators
\begin{eqnarray}
H_{ee}=-\frac{g}{2}\left(F_{+}^{\dagger}F_{+}+F_{-}^{\dagger}F_{-}\right),
\end{eqnarray}
where $ g>0 $ denotes the overall attractive interaction strength. $ H_{ee} $ respects the $ D_{3d}\times\mathbb{T} $ symmetry according to Table.\ref{G}.

Given the model Hamiltonian $ H=H_{e}+H_{ed}+H_{ee} $, we have the partition function $ Z\equiv\text{Tr}\exp\{-H/T\} $ at temperature $T$. To evaluate $ Z $ within the mean field theory, we need to introduce the ferromagnetic and the superconducting order parameters for our system. The ferromagnetic order parameter is defined as the magnetization of the dopants $\bm M\equiv\sum_{\bm r}\langle\bm m(\bm r)\rangle /\mathcal{N}$ where $ \langle\dots\rangle $ denotes the thermodynamic average and $ \mathcal{N} $ denotes the total number of sites. The superconducting order parameters are two complex numbers $ \{\eta_{+},\eta_{-}\} $ and the mean field pairing Hamiltonian can be written as $ H_{\text{P}}^{\dagger} =\eta_{+}F_{+}^{\dagger}+\eta_{-}F_{-}^{\dagger} $.

In terms of the order parameters $ \bm M $ and $ \{\eta_{+},\eta_{-}\} $, the model Hamiltonian $ H $ can be approximated by the following mean field Hamiltonian:
\begin{equation}\label{MF}
H_{\text{MF}}=H_{e}
+I\sum_{i=x,y}S_{i}M_{i}+JS_{z}M_{z}
+H_{\text{P}}^{\dagger}+H_{\text{P}},
\end{equation}
where $ \quad{\bm S}\equiv\sum_{\bm r}\bm s(\bm r) $ denotes the total spin operator of electrons in the system.

Correspondingly, the partition function can be written as $ Z=\int d\eta_{+}d\eta_{-}d^{3}\bm M \exp\{-{F}/{T}\} $, where the mean field free energy is
\begin{equation}
F=\frac{|\bm M|^2}{2\chi_{d}}+\frac{|\eta_{+}|^2 +|\eta_{-}|^2}{2g}-\frac{T}{2}\sum_{{n}}\log[\det {G}_{\text{MF}}^{-1}(i\omega_{n})].
\end{equation}
Here, $ \chi_{d}=\mu_{d}^2/T $ is the magnetic susceptibility of the dopants where $ \mu_{d} $ is the effective magnetic moment of the dopants. $ {G}_{\text{MF}}(i\omega_{n})\equiv (i\omega_{n}-H_{\text{MF}})^{-1} $ is the mean field Gor'kov Green's function with the Matsubara frequency $\omega_{n}$.   

Near the superconducting critical temperature $ T_{c} $, the order parameters $ \eta_{+},\eta_{-},\bm M $ are small in magnitude, and the free energy can be obtained by Taylor expansion on the order parameters as discussed in next section.

{\bf Ginzburg-Landau Analysis}---
The expanded free energy up to the fourth order in $ \eta_{\pm} $ and second order in $ \bm M $ is 
\begin{eqnarray}\label{GL}
F&=&F_{\text{N}}+a(|\eta_{+}|^2 +|\eta_{-}|^2)\\\nonumber
&+&b_{1}(|\eta_{+}|^2 +|\eta_{-}|^2)^2 +b_{2}|\eta_{+}|^2 |\eta_{-}|^2    \\\nonumber
&+& g_{M}M_{z}(|\eta_{+}|^2 -|\eta_{-}|^2)+a_{M}M_{z}^{2}
\end{eqnarray}
where $ F_{\text{N}} $ is the normal phase free energy, and the coefficients $ a,b_{1},b_{2},g_{M},a_{M} $ are expressed in terms of Green's functions (Eqs.(\ref{SMC00})-(\ref{SMC01}) in Section B of the Supplementary Material \cite{SM}). The general form of Eq.(\ref{GL}) can also be easily constructed from the group theory, according to the symmetry properties of the order parameters and their quadratic forms given in Table.\ref{G}.

As the Curie temperature of the dopants should be lower than $ T_c $, in the GL regime we can always assume $ a_{M}>0 $. By minimizing the free energy with respect to $ \bm M $, one obtains the induced magnetization
\begin{eqnarray}\label{GLM}
\bm M =-\frac{g_M}{2a_M}(|\eta_{+}|^2 -|\eta_{-}|^2)\bm z.
\end{eqnarray}
With the expression of $ \bm M $, the free energy is reduced to the function of $ \{\eta_{+},\eta_{-}\} $ only:
\begin{eqnarray}\label{GLSC}
F&=&F_{\text{N}}+a(|\eta_{+}|^2 +|\eta_{-}|^2) \\\nonumber
&+&\left(b_{1}-\frac{g_{M}^2}{4a_{M}}\right)(|\eta_{+}|^2 +|\eta_{-}|^2)^2\\\nonumber
&+&\left(b_{2}+\frac{g_{M}^2}{a_{M}}\right)|\eta_{+}|^2 |\eta_{-}|^2.
\end{eqnarray}

\begin{table}
\centering
\begin{ruledtabular}
\begin{tabular}{c|c|c||c|c} 
IR & electrons & dopants & Superconductivity & Magnetism \\[4pt]
\hline
$A_{1g}$ & $ \bm s^2 $ & $ \bm m^2 $ & $ |\eta_{+}|^2 +|\eta_{-}|^2 $  & $ M_{z}^2 $ \\[4pt]

$A_{2g}$ & $s_{z}$ & $m_{z}$  & $|\eta_{+}|^2 -|\eta_{-}|^2$  & $ M_{z} $\\[4pt]

$E_{g} $ & $\{s_{+},s_{-}\}$ & $\{m_{+},m_{-}\}$ & $ \{\eta_{+}^{*}\eta_{-},\eta_{-}^{*}\eta_{+}\} $  &  $ \{M_{+},M_{-}\} $\\ [4pt]

$E_{u} $ & $ \{F_{+}^{\dagger},F_{-}^{\dagger}\} $ & None & $\{\eta_{+},\eta_{-}\}$  &  None\\ [1pt]
\end{tabular} 
\caption{Irreducible representations (IRs) of operators for electrons/dopants and order parameters for superconductivity/magnetism in $ D_{3d} $. Here, $ \bm s,\bm m $ denote $ \bm s(\bm r),\bm m(\bm r) $ at $ \bm r=\bm 0 $ respectively and $ s_{\pm}\equiv s_{x}\pm is_y $, $ m_{\pm}\equiv m_{x}\pm im_y $, $ M_{\pm}\equiv M_{x}\pm iM_y $.} 
\label{G}
\end{ruledtabular}
\end{table}

Then, we need to minimize the free energy in Eq.(\ref{GLSC}) with respect to $ \{\eta_{+},\eta_{-}\} $. Above $T_c$, $a>0$, the solution $ \eta_{\pm}=0 $ minimizes the free energy, and hence $ \bm M=\bm 0 $ such that the normal phase is paramagnetic. Below $ T_c $, $a<0$, the superconducting phase and $ \bm M $ depend on the sign of  $ b_2 +g_{M}^2/a_{M} $. 

If $ b_2 +g_{M}^2/a_{M} <0 $,  $ |\eta_{+}|=|\eta_{-}|\neq 0 $ minimize the free energy and the system is in the nematic phase  \cite{Fu1, Fu2, Fu3, Fu4}. In the nematic phase, the time-reversal symmetry is preserved and the induced magnetization is zero from Eq.(\ref{GLM}). The three-fold rotational symmetry is found spontaneously broken if we include the sixth order terms in GL free energy as discussed in Ref. \cite{Fu3}. Recent experiments, including Knight shifts and specific heat measurements, provide evidence that this nematic phase is relevant to Cu-doped Bi$_2$Se$_3$ \cite{GQZheng, Fu3, Ando}.

If $ b_2+g_{M}^2/a_{M} >0 $, the chiral phase $ \{\eta_{+},\eta_{-}\}=\Delta\{1,0\} $ or $ \Delta\{0,1\} $ is favored with the pairing amplitude $ \Delta $. The two chiral phases $ \Delta\{1,0\} $ and $ \Delta\{0,1\} $ are time-reversal partners. To be specific, we focus on the phase $ \Delta\{1,0\} $ in the following discussion. In the chiral phase $ \Delta\{1,0\} $, the time-reversal symmetry is spontaneously broken and the total electron spin $ \langle\bm S\rangle ={g_{M}|\Delta|^2\bm z}/{(2J\chi_{d}a_{M})}$ can be finite due to the nonunitary Cooper pairs [$\Gamma_0$, $\Gamma_{xy}$ and $\Gamma_z$ terms in Eq.(\ref{Pairing+})].

The magnetic moments of dopants can couple to the Cooper pairs and result in the magnetic energy of $ E_{d}(\bm M)=\frac{1}{2\chi_{d}}|\bm M|^2+J\langle\bm S\rangle\cdot\bm M $. When $J$ is large enough, the magnetic moments of the dopants are polarized and $\bm M =-{g_M}|\Delta|^2\bm z/({2a_M}) $ as shown in Eq.(\ref{GLM}). Similarly, by minimizing $ E_{d}(\bm M) $, we have $ \bm M=-J\chi_{d}\langle\bm S\rangle $ such that magnetization is proportional to the average spin of the Cooper pairs.  

{\bf Phase Diagram}---
Experimentally, the Cu-doped and Nb-doped Bi$_2$Se$_3$ showed that the superconducting phase can be dramatically changed when the types of dopants are changed.  It was shown that Nb atoms in doped Bi$_2$Se$_3$ have effective magnetic moment of $\mu_d = 1.26\mu_B$ while Cu atoms do not carry magnetic moments \cite{NbTI1}. In this section, by working out the phase diagram for dopants with and without magnetic moments, we show that the magnetic moments of Nb atoms can increase the phase space of the chiral topological phase compared to the Cu-doped case.

As discussed above, the sign of the coefficient $ b_2+g_{M}^2/a_{M} $ in Eq.(\ref{GLSC}) determines whether the system is in the nematic or the chiral phase. The coefficient $ b_2+g_{M}^2/a_{M} $ can be expressed as a function of $\lambda_{a}, \lambda_{b}, \lambda_{c}$ and $J \mu_{d}$, where
\begin{eqnarray}
\lambda_{a}=-\frac{\Gamma_{0}}{\mu}+\frac{\Gamma}{v},\quad\lambda_{b}=\frac{\Gamma_{0}}{\mu}-\frac{\Gamma_{z}}{v_z},\quad\lambda_{c}=\frac{\Gamma_{xy}}{v}.
\end{eqnarray}
The explict form of $ b_2+g_{M}^2/a_{M} $ is given as Eq.(\ref{SMC2}) in Section C of the Supplementary Material \cite{SM}. Fig.\ref{fg2}a depicts the case of Cu-doped Bi$_2$Se$_3$ where $ J\mu_{d}=0 $, which is the same as that shown in Ref. \cite{Fu3}. In Ref. \cite{Fu3}, only the $\Gamma_{0}$ terms are finite and the system is always in the nematic phase (as indicated by the positive infinity along the dashed line in Fig.\ref{fg2}a). In this current work, by considering the more general form of the chiral Cooper pairs, the system can be in the chiral topological phase as found in Nb-doped Bi$_2$Se$_3$. Moreover, by having finite magnetic exchange coupling $J$, the phase space of the chiral topological phase is enlarged. This is simply due to the fact that the Cooper pairs in the chiral topological phase can polarize the magnetic moments of the dopants to further reduce the free energy of the whole system. This is consistent with the experimental finding that Nb-doped samples can possibly be in the chiral topological phase but the Cu-doped samples are possibly in the nematic phase.

{\bf Majorana Nodes and Majorana Arcs of the Chiral Phase}---
In this section we will discuss the band structure of the chiral phase with $ \{\eta_{+},\eta_{-}\}=\Delta\{1,0\} $ and $ \bm M=M\bm z $. This is relevant to the Nb-doped Bi$_2$Se$_3$. Then, the mean field Hamiltonian (\ref{MF}) is simplified to:
\begin{equation}\label{MFC}
H_{\text{MF}}=H_{e}+JMS_{z}
+\Delta(F_{+}^{\dagger}+F_{+}).
\end{equation}

\begin{figure}
\includegraphics[width=3.50in]{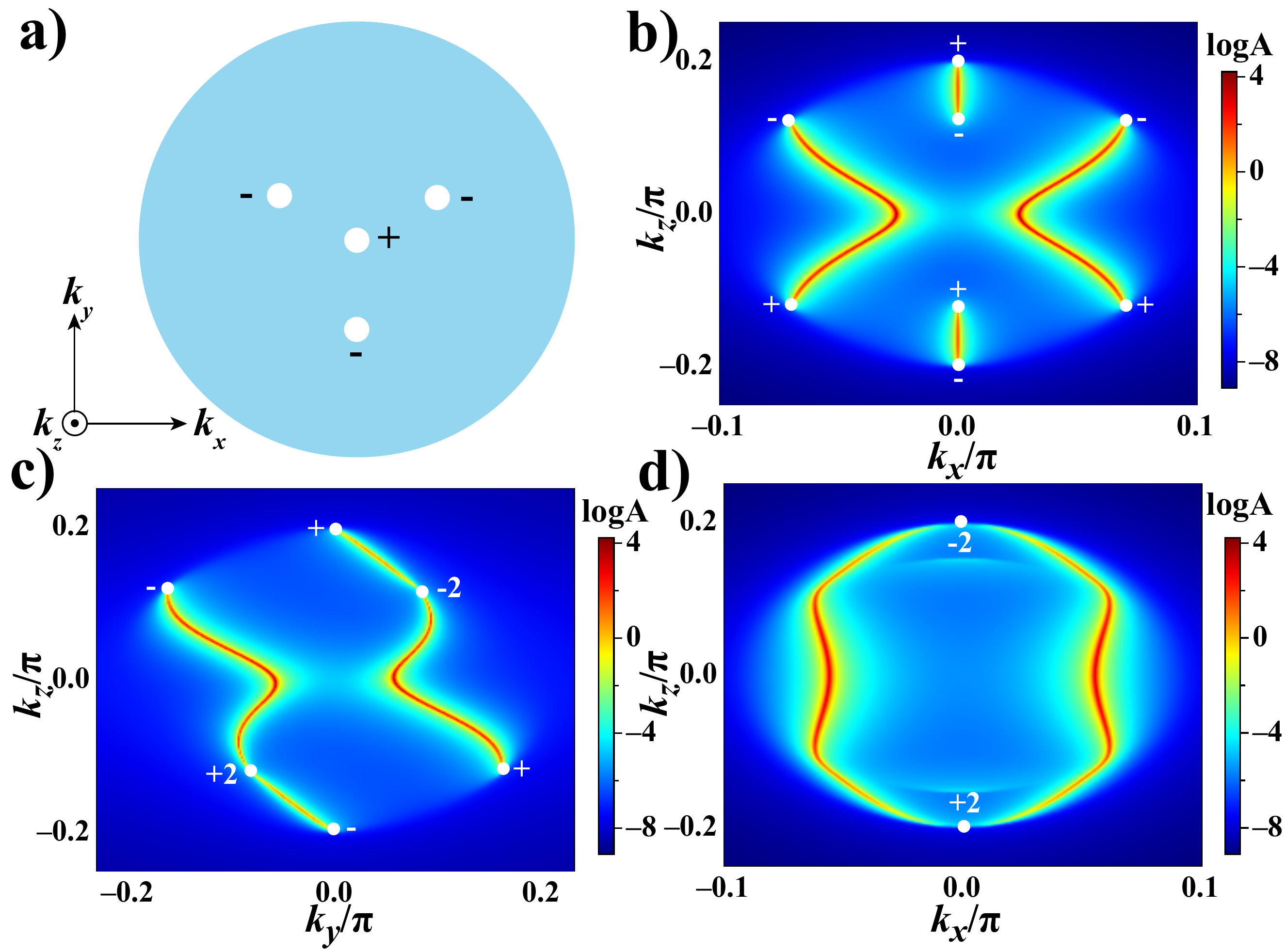}
\caption{a) The schematic top view (from $z$-direction) of nodal points of the chiral superconductor in the Brillouin zone. The four nodes shown are located near the north pole, whose inversion partners (not shown) are the other four nodal points near the south pole. In the general case, the Majorana arcs connecting nodes with opposite chiralities emerge on the surface of superconductors as shown in b) and c). b) Majorana arcs on $xz$-plane. c) Majorana arcs on $yz$-plane. d) The Majorana arcs on $xz$-plane in the special case with $ \Gamma_{0}=\Gamma_{xy}=\Gamma_z =0$ and $\Gamma =0.9$ in Eq.(\ref{Pairing+}) that four nodal points merge into one in the north and south poles respectively. The signs $ \pm $ and numbers $ \pm 2 $ denote the chirality of nodes. In figures b), c) and d) the surface spectral function $ A $ at zero energy is calculated using the Hamiltonian in Eq.(\ref{MFC}), with the parameters $ v=\sqrt{3}v_{z}=5\sqrt{3},m=3,\mu=4,JM=0.2,\Delta =1 $. For b) and c) $ \Gamma_{0}=0.9,\Gamma =0.2,\Gamma_{xy}=0.5,\Gamma_z =0.1 $.}
\label{fg3}
\end{figure}

In general, the mean field Hamiltonian (\ref{MFC}) is nodal with the nodal points lying on the $k_x=k_y=0$ line with finite $k_z$. This is because, when $k_x=k_y=0$, only $\Gamma_0$ and $\Gamma_z$ terms in Eq.(\ref{Pairing+}) can be non-zero. However, these terms pair spin-up electrons while the spin-down electrons cannot be paired. Therefore, there are nodal points in the north and south poles when we have a closed 3D Fermi surface. In general, when $ \lambda_a\lambda_{b}\lambda_{c}\neq 0 $, there are eight nodal points in the energy spectrum. The top view of the four nodal points in the northern hemisphere are shown in Fig.\ref{fg3}a. The four nodal points in the southern hemisphere are inversion partners of the four nodal points in Fig.\ref{fg3}a. Due to the non-degenerate nodal points, the chiral topological phase is indeed a Weyl superconducting phase. The nodal points with opposite chiralities are connected by surface Majorana arcs \cite{Meng,Jay,Yang} as depicted in Figs.\ref{fg3} b to c. Fig.\ref{fg3}b and Fig.\ref{fg3}c depict the Majorana arcs on the $xz$ and $yz$-plane respectively.  The chirality of a node is defined by the Chern number of the Fermi surface enclosing the node. Recently, it was suggested that PrOs$_4$Sb$_{12}$ are chiral superconductors which has similar nodal structures \cite{Fu5}.

In the special case of $\Gamma_0= \Gamma_{xy} = \Gamma_z=0$ but with finite $\Gamma$ in Eq.(\ref{Pairing+}), the four nodal points in the northern hemisphere merge together in the north pole (similarly for the four nodal points in the southern hemisphere).  This results in Weyl nodes with double chirality. The corresponding surface Majorana arcs are depicted in Fig.\ref{fg3}d.

{\bf Conclusion and Discussion}---
In this work, we show that in the chiral phase of Nb doped Bi$_2$Se$_3$, superconductivity can induce ferromagnetic order. In the chiral phase, the system is a Weyl superconductor which possess nodal points and surface Majorana arcs. Interestingly, a recent experiment shows that the three fold rotational symmetry of Nb-doped Bi$_2$Se$_3$ in the superconducting phase is also broken \cite{Lu}. Therefore, further experimental and theoretical investigations are needed to understand this novel superconducting system.

{\bf Acknowledgement} ---
We thank Liang Fu, Yew San Hor, Rolf Lortz and Lu Li for illuminating discussions. The authors thank the support of HKRGC and Croucher Foundation through HKUST3/CRF/13G, 602813, 605512, 16303014 and Croucher Innovation Grant.

\pagebreak

\onecolumngrid

\begin{center}
{\bf \large Supplementary Material for ``Superconductivity-Induced Ferromagnetism and Weyl Superconductivity in Nb-doped Bi$_2$Se$_3$" }
\end{center}

\setcounter{page}{1}
\setcounter{equation}{0}
\setcounter{figure}{0}
\setcounter{table}{0}

\section{A. Symmetry Representation on Two-Orbital Basis}
\renewcommand{\theequation}{A\arabic{equation}}
Microscopically, at site $ \bm r $, the operators $ c_{a/b,\uparrow/\downarrow}^{\dagger}(\bm r) $ create states formed by $p$-orbitals from Bi ($B$) and Se ($S$) atoms with four complex coefficients $ u_{B},v_{B},u_{S},v_{S} $:
\begin{eqnarray}
c_{a/b,\uparrow}^{\dagger}(\bm r)\vert 0\rangle = \vert a/b,\uparrow\rangle &\equiv &
\pm\left[u_{B}\left(\vert B,p_{z},\uparrow\rangle -\vert B',p_{z},\uparrow\rangle\right)+
v_{B}\left(\vert B,p_{+},\downarrow\rangle -\vert B',p_{+},\downarrow\rangle\right)\right]\\\nonumber
&+&
\left[u_{S}\left(\vert S,p_{z},\uparrow\rangle +\vert S',p_{z},\uparrow\rangle\right)+
v_{S}\left(\vert S,p_{+},\downarrow\rangle +\vert S',p_{+},\downarrow\rangle\right)\right]
,\\
c_{a/b,\downarrow}^{\dagger}(\bm r)\vert 0\rangle =\vert a/b,\downarrow\rangle &\equiv &
\pm\left[u_{B}^{*}\left(\vert B,p_{z},\downarrow\rangle -\vert B',p_{z},\downarrow\rangle\right)+
v_{B}^{*}\left(\vert B,p_{-},\uparrow\rangle -\vert B',p_{-},\uparrow\rangle\right)\right]\\\nonumber
&+ &\left[u_{S}^{*}\left(\vert S,p_{z},\downarrow\rangle +\vert S',p_{z},\downarrow\rangle\right)+
v_{S}^{*}\left(\vert S,p_{-},\uparrow\rangle +\vert S',p_{-},\uparrow\rangle\right)\right],
\end{eqnarray}
where $ p_{z},p_{\pm} $ denote $ p $-orbitals with $ z $-component of orbital angular momentum $ 0,\pm 1 $ respectively. The corresponding Bloch basis can be constructed as $ c_{\bm k ls}^{\dagger}\equiv\sum_{\bm r}e^{i\bm k\cdot\bm r}c_{ls}^{\dagger}(\bm r) $, where $ l=a,b $ denotes the orbitals and $ s=\uparrow,\downarrow $ denotes the spin.

The symmetry group $ D_{3d}\times\mathbb{T} $ of the doped Bi$_2$Se$_3$ is generated by the four essential elements: the three-fold rotation $ C_{3z} $, the inversion $ I $, the mirror symmetry $ I_{x} $ and the time-reversal symmetry $ \mathbb{T} $. Under the symmetry operation $ g\in D_{3d}\times\mathbb{T} $, the two-orbital basis $ c_{\bm k}=\{c_{\bm k a\uparrow},c_{\bm k a\downarrow},c_{\bm k b\uparrow},c_{\bm k b\downarrow}\}^{\text{T}} $ transforms as $ g: c_{\bm k}^{\dagger}\to c_{g\bm k}^{\dagger}U(g)$, where
\begin{eqnarray}
U(C_{3z})=\exp(-i\sigma_{z}\pi /3),\quad U(I)=\tau_x,\quad U(I_{x})=-i\sigma_{x},\quad U(\mathbb{T})=i\sigma_{y}\mathbb{K}.
\end{eqnarray}

\section{B. Derivation of Gingzburg-Landau Free Energy From Model Hamiltonian}
\renewcommand{\theequation}{B\arabic{equation}}
The full model Hamiltonian $ H $ of doped Bi$_2$Se$_3$ which respects the symmetry group $ D_{3d}\times\mathbb{T} $ is
\begin{eqnarray}\nonumber
{H}&=&\sum_{\bm k}c^{\dagger}_{\bm k}[ v(k_y\sigma_x -k_x\sigma_y)\tau_z +v_{z}k_{z}\tau_y +m\tau_x -\mu ]c_{\bm k}\\
&+&\sum_{\bm r}\{
I[s_{x}(\bm r)m_{x}(\bm r)+s_{y}(\bm r)m_{y}(\bm r)]
+Js_{z}(\bm r)m_{z}(\bm r)\}-\frac{g}{2}\left(F_{+}^{\dagger}F_{+}+F_{-}^{\dagger}F_{-}\right).
\end{eqnarray}

In order to describe the superconducting phase, we adopt the full Nambu basis $ \{c_{\bm k},c_{-\bm k}^{\dagger\text{T}}\}^{\text{T}} $. On the Nambu basis, the mean field Gor'kov Green's function in the momentum representation reads 
\begin{eqnarray*}
{G}_{\text{MF}}(i\omega_{n},\bm k)&=&
\begin{pmatrix}
i\omega_{n}-H_{0}(\bm k)-I(M_x\sigma_x +M_y\sigma_y)-J M_{z}\sigma_z 
&-\eta_{+}D_{+}(\bm k)-\eta_{-}D_{-}(\bm k)\\
-\eta_{+}^{*}D_{+}^{\dagger}(\bm k)-\eta_{-}^{*}D_{-}^{\dagger}(\bm k)
&i\omega_{n}+H^{\text{T}}_{0}(-\bm k)+I(M_x\sigma_x -M_y\sigma_y)+J M_{z}\sigma_z
\end{pmatrix}^{-1}
\end{eqnarray*}
where $ \omega_{n}=(2n+1)\pi T (n\in\mathbb{Z}) $ is the fermionic Matsubara frequency, $ H_{0}(\bm k)= v(k_y\sigma_x -k_x\sigma_y)\tau_z +v_{z}k_{z}\tau_y +m\tau_x -\mu $ is the normal phase Hamiltonian of doped Bi$_2$Se$_3$, and $D_{\pm}(\bm k)=(\Gamma_{0}\sigma_{\pm}\tau_{y}+\Gamma k_{\pm}\sigma_{z}\tau_{x}
\pm i\Gamma_{xy}k_{\mp}\sigma_{\mp}-\Gamma_{z}k_{z}\sigma_{\pm})i\sigma_{y}$ denotes the pairing matrices corresponding to $ F_{\pm}^{\dagger} $ respectively. With $ {G}_{\text{MF}}(i\omega_{n},\bm k) $, the mean field free energy at temperature $T$ reads
$ F={|\bm M|^2}/{2\chi_{d}}+({|\eta_{+}|^2 +|\eta_{-}|^2})/{2g}-\frac{1}{2}T\sum_{{n},\bm k}\log[\det {G}_{\text{MF}}^{-1}(i\omega_{n},\bm k)] $. The next step is to expand $ F $ in terms of superconducting and magnetic order parameters, equivalently the expansion of $ {G}_{\text{MF}}(i\omega_{n},\bm k) $ in terms of $ \{\eta_{+},\eta_{-}\} $ and $ \bm M $. We write the inverse Gor'kov Green's function as
${G}_{\text{MF}}^{-1}={G}^{-1}_{0}(1-\mathcal{D}-\mathcal{M})$
where $ {G}_{0}=\text{diag}[G_e ,G_h] $ is the normal Green's function. Here $ G_e (i\omega_n ,\bm k)=[i\omega_n -H_{0}(\bm k)]^{-1},G_h (i\omega_n)=[i\omega_n +H_{0}^{\text{T}}(-\bm k)]^{-1} $ are Green's functions for electrons and holes which can be explicitly written as
\begin{eqnarray}
G_{e}(i\omega_{n},\bm k)=\frac{i\omega_{n}-H_{0}(\bm k)}{(i\omega_{n}+\mu)^2-\varepsilon(\bm k)^2},\quad G_{h}(i\omega_{n},\bm k)=\frac{i\omega_{n}+H_{0}^{*}(-\bm k)}{(i\omega_{n}-\mu)^2-\varepsilon(\bm k)^2},
\end{eqnarray}
with $ \varepsilon(\bm k)=\sqrt{m^2 +v^2(k_{x}^2+k_{y}^2)+ v_{z}^{2}k_{z}^2} $. The matrices $\mathcal{D}, \mathcal{M}$ are
\begin{eqnarray}\label{GFS}
\mathcal{D}(i\omega_{n},\bm k)&=&
\begin{pmatrix}
0&G_e (i\omega_n ,\bm k)[\eta_{+}D_{+}(\bm k)+\eta_{-}D_{-}(\bm k)]\\
G_h (i\omega_n ,\bm k)[\eta_{+}^{*}D_{+}^{\dagger}(\bm k)-\eta_{-}^{*}D_{-}^{\dagger}(\bm k)]&0
\end{pmatrix}\\
\mathcal{M}(i\omega_{n},\bm k)&=&
\begin{pmatrix}
G_e (i\omega_n ,\bm k)[I(M_x\sigma_x +M_y\sigma_y)+J M_{z}\sigma_z]&0\\
0&-G_h (i\omega_n ,\bm k)[I(M_x\sigma_x -M_y\sigma_y)+J M_{z}\sigma_z]
\end{pmatrix}.
\end{eqnarray}

Near superconducting critical temperature $T_c$, $ |\eta_{\pm}|,|\bm M| $ are small in magnitude, and hence $ \mathcal{D},\mathcal{M} $ are small in norm. Expanded in terms of $ \mathcal{D},\mathcal{M} $ up to the fourth order in $ \mathcal{D} $ and second order in $ \mathcal{M} $, the mean field contribution to the free energy reads
\begin{eqnarray}
-\log\det {G}_{\text{MF}}^{-1}&=&-\text{tr}\log[{G}^{-1}_{0}(1-\mathcal{D}-\mathcal{M})]
=-\text{tr}\log{G}^{-1}_{0}-\text{tr}\log(1-\mathcal{D}-\mathcal{M})\\ \nonumber
&=&-\text{tr}\log{G}^{-1}_{0}
+\frac{1}{2}\text{tr}(\mathcal{D}^2)
+\frac{1}{2}\text{tr}(\mathcal{D}^4)
+\text{tr}(\mathcal{M} \mathcal{D}^2)
+\frac{1}{2}\text{tr}\mathcal{M}
+\frac{1}{2}\text{tr}(\mathcal{M}^2)+\dots
\end{eqnarray}

Thus expanded in terms of $ \eta_{\pm},\bm M $ up to the fourth order in $ \eta_{\pm} $ and second order in $ \bm M $, the total free energy reads
\begin{eqnarray}
F&=&F_{\text{N}}
+\sum_{\alpha ,\beta =\pm}a_{\alpha\beta}\eta_{\alpha}^{*}\eta_{\beta}
+\sum_{\alpha ,\beta ,\alpha' ,\beta' =\pm}
b_{\alpha\beta\alpha'\beta'}
\eta_{\alpha}^{*}\eta_{\beta}\eta_{\alpha'}^{*}\eta_{\beta'}+\sum_{\alpha ,\beta =\pm}\sum_{i=x,y,z}g_{\alpha\beta i}\eta_{\alpha}^{*}\eta_{\beta}M_{i}
+\sum_{i,j=x,y,z}\tau_{ij}M_{i}M_{j}
\end{eqnarray}
where $ F_{\text{N}}=-\frac{1}{2}T\log\det G_{0}^{-1} $ is the normal phase free energy and the GL coefficients are
\begin{eqnarray}\label{SMC00}
a_{\alpha\beta}&=&
\frac{\delta_{\alpha\beta}}{2g}
+\frac{1}{2}T\sum_{{n} ,\bm k}\text{tr}[
{D}^{\dagger}_{\alpha}(\bm k)G_e(i\omega_n ,\bm k)
{D}_{\beta}(\bm k) G_h(i\omega_n ,\bm k)]\\
b_{\alpha\beta\alpha'\beta'}&=&\frac{1}{4}T\sum_{{n} ,\bm k}\text{tr}[
{D}^{\dagger}_{\alpha}(\bm k)G_e(i\omega_n ,\bm k)
{D}_{\beta}(\bm k) G_h(i\omega_n ,\bm k)
{D}^{\dagger}_{\alpha'}(\bm k)G_e(i\omega_n ,\bm k)
{D}_{\beta'}(\bm k) G_h(i\omega_n ,\bm k)]\\
g_{\alpha\beta i}&=&\frac{1}{2}{J}_{i}T\sum_{{n} ,\bm k}\text{tr}\{
{D}^{\dagger}_{\alpha}(\bm k)G_e(i\omega_n ,\bm k)[\sigma_{i}
G_e(i\omega_n ,\bm k){D}_{\beta}(\bm k)-
{D}_{\beta}(\bm k)G_h(i\omega_n ,\bm k)
\sigma_{i}^{\text{T}}]G_h(i\omega_n ,\bm k)\}\\\label{SMC01}
\tau_{ij}&=&\frac{\delta_{ij}}{2\chi_{d}}+\frac{1}{4}{J}_{i}^2 T\sum_{{n} ,\bm k}\text{tr}\{[
G_{e}(i\omega_n ,\bm k)\sigma_{i}G_{e}(i\omega_n ,\bm k)\sigma_{j}+
G_{h}(i\omega_n ,\bm k)\sigma_{i}^{\text{T}}G_{h}(i\omega_n ,\bm k)\sigma_{j}^{\text{T}}]\}
\end{eqnarray}
where $ {J}_{x}={J}_{y}=I,\quad {J}_z =J $.

One can apply $ D_{3d}\times\mathbb{T} $ to $ D_{\pm}(\bm k), G_{e/h}(i\omega_n ,\bm k) $ and obtains the symmetry properties of the GL coefficients
\begin{equation*}
a_{\alpha\beta}=a_{-\alpha, -\beta}^{*}=a_{-\alpha, -\beta}=e^{-2i\pi (\alpha -\beta)/3}a_{\alpha\beta},\quad
b_{\alpha\beta\alpha'\beta'}=b_{-\alpha, -\beta ,-\alpha' ,-\beta'}^{*}=b_{-\alpha, -\beta ,-\alpha' ,-\beta'}=e^{-2i\pi (\alpha +\alpha' -\beta -\beta')/3}b_{\alpha\beta\alpha'\beta'},
\end{equation*}
\begin{equation*}
g_{\alpha\beta x}=g_{-{\alpha},-\beta , x},\quad
g_{\alpha\beta y}=-g_{-{\alpha},-\beta , y},\quad
g_{\alpha\beta z}=-g_{-{\alpha},-\beta , z},\quad
g_{\alpha\beta i}=-g_{-{\alpha},-\beta , i}^{*},
\end{equation*}
\begin{equation*}
g_{\alpha\beta z}=e^{-2i\pi (\alpha -\beta)/3}g_{\alpha\beta z},\quad
g_{\alpha\beta x}\pm ig_{\alpha\beta y}=e^{-{2i\pi}(\alpha -\beta \pm 1)/3}(g_{\alpha\beta x}\pm ig_{\alpha\beta y}),\quad
\tau_{ij}=\tau_{i}\delta_{ij},\quad \tau_{x}=\tau_{y}.
\end{equation*}
The symmetry constraints of the coefficients guarantee the following GL free energy as shown in the maintext
\begin{eqnarray}
F=F_{\text{N}}+a(|\eta_{+}|^2 +|\eta_{-}|^2)
+b_{1}(|\eta_{+}|^2 +|\eta_{-}|^2)^2 +b_{2}|\eta_{+}|^2 |\eta_{-}|^2 
+g_{M}M_{z}(|\eta_{+}|^2 -|\eta_{-}|^2)+a_{M}M_{z}^{2}
\end{eqnarray}
where $ a=a_{++},\quad b_{1}=b_{++++},\quad b_{2}=2(b_{++--}+b_{+--+}-b_{++++}),\quad g_{M}=g_{++z},\quad a_{M}=\tau_{zz} $.

At last we will calculate the expectation value of the total electron spin $\langle\bm S\rangle\equiv\sum_{\bm r}\langle \bm s(\bm r)\rangle$. In order to do that, a virtual Zeeman field $ \bm B $ coupled with the total electron spin $\bm S $ is introduced to the model Hamiltonian $ H $
\begin{eqnarray}\nonumber
{H}_{\bm B}&=&H+\bm B\cdot\bm S
\end{eqnarray}
and the Gor'kov Green's function is modified correspondingly ${G}_{\bm B}^{-1}={G}^{-1}_{0}(1-\mathcal{D}-\mathcal{M}-\mathcal{B})$
where 
\begin{eqnarray}
\mathcal{B}(i\omega_{n},\bm k)&=&\bm B\cdot
\begin{pmatrix}
G_e (i\omega_n ,\bm k)\bm\sigma &0\\
0&-G_h (i\omega_n ,\bm k)\bm\sigma^{\text{T}}
\end{pmatrix}.
\end{eqnarray}

So we find the expansion of the mean field contribution to the free energy
\begin{eqnarray}
-\log\det {G}^{-1}_{\bm B}=-\text{tr}\log[{G}^{-1}_{0}(1-\mathcal{D}-\mathcal{M})]+\text{tr}(\mathcal{M}\mathcal{B})+\text{tr}(\mathcal{D}^2 \mathcal{B})+\dots
\end{eqnarray}
and hence of the total modified free energy
\begin{eqnarray}\nonumber
F_{\bm B}=F+\frac{g_{M}}{J}(|\eta_{+}|^2 -|\eta_{-}|^2)B_{z}+\frac{2a_{M}-\chi_{d}^{-1}}{J}M_{z}B_{z}
\end{eqnarray}
up to the first order in $ \bm B $.

Thus $\langle\bm S\rangle$ can be calculated through the derivatives of the modified free energy:
\begin{eqnarray}\label{SMC1}
\langle\bm S\rangle =\left.\frac{\partial F_{\bm B}}{\partial\bm B}\right|_{\bm B=\bm 0}=\frac{1}{J}\left[g_{M}(|\eta_{+}|^2 -|\eta_{-}|^2)+(2a_{M}-\chi_{d}^{-1})M_{z}\right]\bm z =\frac{g_{M}}{2J\chi_{d}a_{M}}(|\eta_{+}|^2 -|\eta_{-}|^2)\bm z.
\end{eqnarray}
One immediately finds that the induced magnetization is proportional to the total electron spin $ \bm M=-J\chi_{d}\langle\bm S\rangle $.

\section{C. Phase Diagrams and Band Structure}
\renewcommand{\theequation}{C\arabic{equation}}
From the expressions of the GL coefficients and Green's functions, $ b_{2},g_{M},a_{M} $ can be worked out as
\begin{eqnarray}
b_2&=&\frac{7\zeta(3)N(0)(\lambda_{c}\sqrt{\mu^2 -m^2})^{4}}{120(\pi T_{c})^2}\left[\left(\frac{\lambda_a}{\lambda_c}\right)^4+4\left(\frac{\lambda_b}{\lambda_c}\right)^2-8\left(1+\frac{\lambda_a^2}{\lambda_c^2}\right)-2\frac{\lambda_b^2}{\lambda_c^2}\left(\frac{\lambda_a^2}{\lambda_c^2}-2\right)-3\frac{\lambda_b^4}{\lambda_c^4}\right],\\
g_M&=&-\pi\log\left(\frac{2e^{\gamma_{E}}\omega_c}{\pi T_c}\right)JN'\left(0\right) 
(\mu^2 -m^2)\left[\frac{m}{\mu}
\left(-\frac{2}{3}\lambda_c^2+\frac{2}{3}\lambda_b^2\right)
+\left(1-\frac{m}{\mu}\right)
\left(\frac{1}{5}\lambda_b^2-\frac{2}{15}\lambda_c^2-\frac{2}{15}\lambda_a\lambda_b\right)\right],\\
a_{M}&=&\frac{1}{2}\left[\frac{T_{c}}{\mu_{d}^2}-\frac{1}{3}J^2N(0)\mu_{B}^2\left(1+\frac{2m^2}{\mu^2}\right)\right],
\end{eqnarray}
where $ N(0) $ is the density of states per spin in the conduction band, $ \omega_c $ is the cutoff energy and $ \gamma_{E}=0.57721\dots $ is the Euler gamma constant. In performing the calculations we assume the superconductivity happens within the energy shell $ |\xi|\leq\omega_c $.

With the GL coefficients given above, one obtains the quantity determining the phase diagrams
\begin{align}\label{SMC2}
b_2+\frac{g_M^2}{a_M}&=\frac{7\zeta(3)N(0)(\lambda_{c}\sqrt{\mu^2 -m^2})^{4}}{120(\pi T_{c})^2}\left\{\left(\frac{\lambda_a}{\lambda_c}\right)^4+4\left(\frac{\lambda_b}{\lambda_c}\right)^2-8\left(1+\frac{\lambda_a^2}{\lambda_c^2}\right)-2\frac{\lambda_b^2}{\lambda_c^2}\left(\frac{\lambda_a^2}{\lambda_c^2}-2\right)-3\frac{\lambda_b^4}{\lambda_c^4}\right.\\\nonumber
&\left.+L(J\mu_d)\left[\frac{m}{\mu}\left(-\frac{2}{3}+\frac{\lambda_b^2}{3\lambda_c^2}\right)+\left(1-\frac{m}{\mu}\right)\left(\frac{\lambda_b^2}{5\lambda_c^2}-\frac{2}{15}-\frac{2\lambda_a\lambda_b}{15\lambda_c^2}\right)\right]^2\right\},
\end{align}
where 
\begin{eqnarray}
L(x)=\frac{A_{0}x^2}{1-C x^2},\quad A_{0}= \frac{120\pi^4 T_{c}}{7\zeta(3)N(0)}\left[N'(0)\log\left(\frac{2e^{\gamma_{E}}\omega_c}{\pi T_c}\right)\right]^2,\quad C=\frac{2N(0)\mu_{B}^2}{3T_{c}}\left(1+\frac{2m^2}{\mu^2}\right).
\end{eqnarray}
The function $ L(x) $ is a dimensionless monotonic function of $ x $ in the range $ 0<x<C^{-\frac{1}{2}} $. For Cu-doped Bi$_2$Se$_3$, $ J\mu_{d}=0 $ hence $ L=0 $, and the phase diagram is obtained as shown in Fig.\ref{fg2}a. For Nb-doped Bi$_2$Se$_3$, we choose $ m/\mu=2/3,L=30 $ to work out the phase diagram Fig.\ref{fg2}b.

To obtain the band structure we construct the following tight-binding model of the chiral superconductor
\begin{eqnarray}
H_{0}(\bm k)&=&\frac{2v}{3}
\begin{pmatrix}
0& iS_{-}(\bm k)\\
-iS_{+}(\bm k)& 0
\end{pmatrix}
\tau_z +v_{z}\sin k_{z}\tau_y +m\tau_x -tC(\bm k)+3t-\mu\\
D_{\pm}(\bm k)&=&\left[\Gamma_{0}\sigma_{\pm}\tau_{y}+\frac{2}{3}\Gamma S_{\pm}(\bm k)\sigma_{z}\tau_{x}
\pm \frac{2i}{3}\Gamma_{xy}S_{\mp}(\bm k)\sigma_{\mp}-\Gamma_{z}\sin k_{z}\sigma_{\pm}\right]i\sigma_{y}\\
H_{\text{MF}}(\bm k)&=&
\begin{pmatrix}
H_{0}(\bm k)+JM\sigma_{z}\tau_0 & \Delta D_{+}(\bm k)\\
\Delta D_{+}^{\dagger}(\bm k) & -H_{0}^{*}(-\bm k)-JM\sigma_{z}\tau_0
\end{pmatrix}.
\end{eqnarray}
The special functions are defined as
\begin{eqnarray}
S_{\pm}(\bm k)\equiv\sum_{j=0}^{2}\omega^{j}\sin(\bm k\cdot\bm R_{j})\simeq\frac{3}{2}(k_{x}\pm ik_{y}),\quad C(\bm k)\equiv\sum_{j=0}^{2}\cos(\bm k\cdot\bm R_{j})\simeq 3-\frac{3}{4}|\bm k|^2,\quad (|\bm k|\to 0)
\end{eqnarray}
where $ \omega =e^{2\pi i/3},\bm R_{0}=\bm x,\bm R_{1}=({-\bm x+\sqrt{3}\bm y})/{2},\bm R_{2}=({-\bm x+\sqrt{3}\bm y})/{2} $.

Importantly, the term $ -tC(\bm k)+3t $ is introduced. Without this term, the $ \Gamma $ point and two $ K $ points will be at the same energy level accidentally, which does not match the experimental results of ARPES. In Fig.\ref{fg3}b, c and d of the maintext we set $ t=1 $.

\end{document}